\documentclass{Poitiers}

\usepackage{graphicx}

\def\ppages{1--10}

\theoremstyle{plain}

\theoremstyle{definition}
\newtheorem{definition}{Definition}

\title[SIMULATIONS OF FIRST PASSAGE TIME]
 {First Passage Time for Multivariate Jump-diffusion Stochastic Models
With Applications in Finance }

\email{rmelnik@wlu.ca}



\author[Di Zhang and Roderick V.N. Melnik]{}

\begin{document}
\maketitle

\centerline{\scshape Di Zhang and Roderick V.N. Melnik}
{\footnotesize \centerline{}
 \centerline{Mathematical Modelling and Computational Sciences}
 \centerline{Wilfrid Laurier University}
 \centerline{Waterloo, ON, Canada N2L 3C5}
}

\begin{quote}{\normalfont\fontsize{8}{10}\selectfont
{\bfseries Abstract.} The ``first passage-time'' problem is an
important problem with a wide range of applications in mathematics,
physics, biology and finance. Mathematically, such a problem can be
reduced to estimating the probability of a (stochastic) process
first to reach a critical level or threshold. While in other areas
of applications the FPT problem can often be solved analytically, in
finance we usually have to resort to the application of numerical
procedures, in particular when we deal with jump-diffusion
stochastic processes (JDP). In this paper, we develop a
Monte-Carlo-based methodology for the solution of the FPT problem in
the context of a multivariate jump-diffusion stochastic process. The
developed methodology is tested by using different parameters, the
simulation results indicate that the developed methodology is much
more efficient than the conventional Monte Carlo method, which
establishes itself as an efficient tool for further practical
applications, such as the analysis of default correlation and
predicting barrier options in finance.\par}
\end{quote}

\section{Introduction}

In a jump-diffusion process (JDP), the dynamics of underlying
process have two random components: a continuous diffusion component
and a discontinuous jump component \cite{Zhou:2001:jump}, in which
the jump component can be explained as a sudden drop of process's
value. The first passage time (FPT) problem for jump-diffusion
processes has Attracted attention of researchers in such diverse
fields as queuing networks \cite{Gorov:1985}, computer vision
\cite{Zhu:1999}, target recognition \cite{Miller:1997}. In the
financial world, many problems also require the information on the
first passage time of a stochastic process, for example, in modeling
credit risk and valuing defaultable securities
\cite{Zhou:2001:jump}, or in predicting barrier options
\cite{Metwally:2002}. Furthermore, it is now generally accepted that
the geometric Brownian motion model for market behavior may produce
misleading results in \cite{Zhou:2001:jump,Atiya:2005}, such as
mismatching credit spreads on corporate bonds, underlying derivative
prices. Jump-diffusion processes have established themselves as a
sound alternative to the geometric Brownian motion model.

However, if we consider jumps in the process, except for very basic
process types where closed form solutions are available, when the
jump sizes are doubly exponential or exponentially distributed
\cite{Kou:2003}, or when the jumps can have only nonnegative values
(assuming that the crossing boundary is below the process starting
value) \cite{Blake:1973}, where closed form solutions are available,
for most problems we can only resort to the numerical procedures.

Monte Carlo simulation is a very promising candidate in dealing with
FPT problems. However, in conventional Monte Carlo method, in order
to avoid discretization bias \cite{Platen:2003}, we need to
discretize the time horizon into small enough intervals, and to
evaluate the process at each discretized time that is very
time-consuming. Recently, Atiya and Metwally
\cite{Metwally:2002,Atiya:2005} have developed a fast Monte
Carlo-type numerical method to solve the FPT problem for
jump-diffusion process.

In many financial problems we have to deal with multiple processes
in practice and consider their jointly crossing the critical level,
so it is very useful to develop fast numerical procedure for
multivariate jump-diffusion process. In this contribution, we extend
the fast Monte Carlo-type numerical methodology to the more general
case that covers affine multivariate processes jump-diffusion. The
developed methodology can be easily extended to other financial
applications and areas where FPT problem arises.

The article is organized as follows: section \ref{Model} describes
our mathematical model. The algorithms are presented in section
\ref{Methodology} and simulation results are given in section
\ref{Simulations}. Concluding remarks are given section
\ref{Conclusion}.

\section{Mathematical Model}
\label{Model} In this section, first, we give brief discussion on
the affine jump-diffusion model, then we deduce the first passage
time distribution under multivariate jump-diffusion process. At
last, we consider the issue of kernel estimator.

\subsection{Affine jump-diffusion}
\label{subsection:AJD}

Affine jump-diffusion is a jump-diffusion process for which the
drift vector, ``instantaneous'' covariance matrix and jump
intensities all have affine dependence on the state vector.

Let us consider a complete probability space $(\Omega,F,P)$ and an
information filtration $(F_t)$, and suppose that $X$ is a Markov
process in some state space $D\subset\mathbb{R}^n$, solving the
stochastic differential equation \cite{Duffie:2000}
\begin{equation}
  dX_t=\mu(X_t)dt+\sigma(X_t)dW_t+dZ_t,
  \label{AJD:DiffEq}
\end{equation}
where $W$ is an $(F_t)$-standard Brownian motion in $\mathbb{R}^n$;
$\mu:D\rightarrow\mathbb{R}^n$,
$\sigma:D\rightarrow\mathbb{R}^{n\times n}$, and $Z$ is a pure jump
process whose jumps have a fixed probability distribution $\nu$ on
$\mathbb{R}^n$ and arrive with intensity $\{\lambda(X_t):t\ge 0\}$,
for some $\lambda:D\rightarrow[0,\infty)$.

\begin{definition}
The above model is an affine model if \cite{Duffie:2000}:
\begin{eqnarray}
  & & \mu(X_t,t) = K_0 + K_1 X_t\nonumber\\
  & & (\sigma(X_t,t)\sigma(X_t,t)^\top)_{ij} = (H_0)_{ij}+(H_1)_{ij}X_j\nonumber\\
  & & \lambda(X_t) = l_0+l_1\cdot X_t,
  \label{Eq:AJD:terms}
\end{eqnarray}
where $K=(K_0,K_1)\in\mathbb{R}^n\times\mathbb{R}^{n\times n}$,
$H=(H_0,H_1)\in\mathbb{R}^{n\times n}\times\mathbb{R}^{n\times
n\times n}$, $l=(l_0,l_1)\in\mathbb{R}^n\times\mathbb{R}^{n\times
n}$ and the ``jump transform'' (determines the jump-size
distribution) $\psi(c)=\int_{\mathbb{R}^n}\exp(c,z)d\nu(z)$, for
$c\in\mathbb{C}^n$, is known whenever the integral is well defined.
The ``coefficients'' $(K,H,l,\psi)$ of $X$ completely determine its
distribution.
\end{definition}

\subsection{First passage time distribution}
\label{subsection:FPTD}

Now, we will consider the first passage time distribution in the
context of multiple processes. In order to obtain a computable
multi-dimensional solutions of FPT distribution, we need to simplify
Eq. (\ref{AJD:DiffEq}) and (\ref{Eq:AJD:terms})based on the
following assumptions:
\begin{enumerate}
  \item Each $W_t$ in Eq. (\ref{AJD:DiffEq}) is independent;
  \item $K_1=0$, $H_1=0$ and $l_1=0$ that means the drift term, the diffusion
  process (Brownian motion) and the arrival intensity are independent with state
  vector $X$;
  \item The distribution of jump-size $Z_t$ is also independent with $X$.
\end{enumerate}

In this scenario, we can rewrite Eq. (\ref{AJD:DiffEq}) as
\begin{equation}
  dX_t=\mu dt+\sigma dW_t+dZ_t,
  \label{JDP:multi}
\end{equation}
where
\[
  \mu = K_0,\;\sigma\sigma^\top = H_0,\;\lambda = l_0.
\]

Atiya \textit{et al.} \cite{Atiya:2005} have deduced one-dimensional
first passage time distribution in time horizon $[0,T]$. In order to
judge multiple processes, from Eq. (\ref{JDP:multi}), we obtain
multi-dimensional formulas and simplify them into computable
formulas. We will highlight the main steps of our procedure below.

As defined, the multiple processes $X$ can be written as
$X=[X_1,X_2,\dots]^\top$. Let us consider one of its components, a
sub-process $X_i$ that satisfies the following stochastic
differential equation:
\begin{eqnarray}
  dX_i & = & \mu_{i}dt+\sum_{j}\sigma_{ij}dW_j+dZ_i\nonumber\\
       & = & \mu_{i}dt+\sigma_{i}dW_i+dZ_i,
  \label{JDP:one}
\end{eqnarray}
where $W_i$ is also a standard Brownian motion and $\sigma_{i}$ is:
\[
  \sigma_{i}=\sqrt{\sum_{j}\sigma_{ij}^2}.
\]

We assume that in the interval $[0,T]$, $M_i$ times of jumps happen
for $X_i$. Let the jump instants be $T_{1}, T_{2},\cdots,T_{M_i}$.
Let $T_{0}=0$ and $T_{M_i+1}=T$. $\tau_j$ equals interjump times,
which is $T_{j}-T_{j-1}$. Following the notation of
\cite{Atiya:2005}, let $X_{i}(T_{j}^{-})$ be the process value
immediately before the $j$th jump, and $X_{i}(T_{j}^{+})$ be the
process value immediately after the $j$th jump. The jump-size is
then $X_{i}(T_{j}^{+})-X_{i}(T_{j}^{-})$, and we can use this
jump-size to generate $X_{i}(T_{j}^{+})$ sequentially.

If we define $A_i(t)$ as the event that process crossed the
threshold $D_i(t)$ for the first time in the interval $[t,t+dt]$, we
have
\begin{equation}
  g_{ij}(t)=p(A_i(t)\in dt|X_i(T_{j-1}^{+}),X_i(T_{j}^{-})).
\end{equation}

If we only consider one interval $[T_{j-1},T_{j}]$, we can obtain
\cite{Feller:1968,Rogers:1994}
\begin{eqnarray}
  g_{ij}(t) & = &
  \frac{X_i(T_{j-1}^{+})-D_{i}(t)}{2y_i\pi\sigma_{i}^{2}}(t-T_{j-1})^{-\frac{3}{2}}(T_{j}-t)^{-\frac{1}{2}}\nonumber\\
  & & *\exp\left(-\frac{[X_i(T_{j}^{-})-D_{i}(t)-\mu_{i}(T_{j}-t)]^{2}}{2(T_{j}-t)\sigma_{i}^{2}}\right)\nonumber\\
  & & *\exp\left(-\frac{[X_i(T_{j-1}^{+})-D_{i}(t)+\mu_{i}(t-T_{j-1})]^{2}}{2(t-T_{j-1})\sigma_{i}^{2}}\right),
  \label{FPTD:condition}
\end{eqnarray}
where
\[
  y_i=\frac{1}{\sigma_{i}\sqrt{2\pi\tau_{j}}}
    \exp\left(-\frac{[X_i(T_{j-1}^{+})-X_i(T_{j}^{-})+\mu_{i}\tau_{j}]^{2}}{2\tau_{j}\sigma_{i}^{2}}\right).
\]

After getting result in one interval, we combine the results to
obtain the density for the whole interval $[0,T]$. Let $B(s)$ be a
Brownian bridge in the interval $[T_{j-1},T_{j}]$, with
$B(T_{j-1}^{+})=X_i(T_{j-1}^{+})$, $B(T_{j}^{-})=X_i(T_{j}^{-})$,
the probability that the minimum of $B(s_i)$ is always above the
boundary level is \cite{Karatzas:1991}
\begin{eqnarray}
  P_{ij} & = & P\left(\inf_{T_{j-1}\leq s_i\leq T_{j}}B(s_i)>D_{i}(t)|B(T_{j-1}^{+})=X_i(T_{j-1}^{+}),B(T_{j}^{-})=X_i(T_{j}^{-})\right)\nonumber\\
   & = & \left\{
     \begin{array}{ll}
       1-\exp\left(-\frac{2[X_i(T_{j-1}^{+})-D_{i}(t)][X_i(T_{j}^{-})-D_{i}(t)]}{\tau_{j}\sigma_{i}^{2}}\right), & \mathrm{if}\;X_i(T_{j}^{-})>D_{i}(t),\\
       0, & \mathrm{otherwise}.
     \end{array}\right.
   \label{BM:default}
\end{eqnarray}

Then $B(s_i)$ is below the threshold level, which means the default
happens or already happened, and its probability is $1-P_{ij}$. Then
let $L(s_i)\equiv L_i$ denote the index of the interjump period in
which the time $s_i$ falls in $[T_{L_i-1},T_{L_i}]$. Also let $I_i$
represent the index of the first jump, which happened in simulated
jump instant.
\begin{eqnarray}
  I_i & = & \min(j:X_i(T_{k}^{-})>D_{i}(t);k=1,\ldots,j,\;\mathrm{and}\nonumber\\
    &   & \;\;\;\;\;\;\;\;\;\;\;\:X_i(T_{k}^{+})>D_{i}(t);k=1,\ldots,j-1,\;\mathrm{and}
          \;X_i(T_{j}^{+})\leq D_{i}(t)).
  \label{index_first_jump}
\end{eqnarray}

If no such $I_i$ exists, we set $I_i=0$.

Then we get the probability of the interval $[0,T]$

\begin{eqnarray}
  & & P(A_i(s_i)\in ds|X_i(T_{j-1}^{+}),X_i(T_{j}^{-}),j=1,\ldots,M_{i}+1)\nonumber\\
   & = & \left\{
     \begin{array}{ll}
       g_{iL_i}(s_i)\prod_{k=1}^{L_i-1}P_{ik} & \mathrm{if}\;L_i<I_i\;\mathrm{or}\;I_i=0,\\
       g_{iL_i}(s_i)\prod_{k=1}^{L_i-1}P_{ik}+\prod_{k=1}^{L_i}P_{ik}\delta(s_i-T_{I_i}) & \mathrm{if}\;L_i=I_i,\\
       0 & \mathrm{if}\;L_i>I_i,
     \end{array}\right.
\end{eqnarray}
where $\delta$ is the Dirac's delta function.

\subsection{The kernel estimator}
\label{subsection:estimation}

For each $X_i$, after generating series of first passage times
$s_i$, we use a kernel density estimator with a Gaussian kernel to
estimate the first passage time density (FPTD) $f$. As described in
\cite{Atiya:2005}, the kernel density estimator is based on
centering a kernel function of a bandwidth as follows:
\begin{equation}
  \widehat{f}=\frac{1}{N}\sum_{i=1}^{N}K(h,t-s_{i}),
\end{equation}
where
\begin{equation}
  K(h,t-s_{i})=\frac{1}{\sqrt{\pi/2}h}\exp\left(-\frac{(t-s_{i})^{2}}{h^2/2}\right).
\end{equation}

The optimal bandwidth in the kernel function $K$ can be calculated
as \cite{Silverman:1986}:
\begin{equation}
  h_{opt}=\left(2N\sqrt{\pi}\int_{-\infty}^{\infty}(f_{t}'')^{2}dt\right)^{-0.2},
  \label{estamate:hopt}
\end{equation}
where $N$ is the number of generated points and $f_{t}$ is true
density. Here we use the approximation for the distribution as a
gamma distribution as proposed in \cite{Atiya:2005}:
\begin{equation}
  f_{t}=\frac{\alpha^{\beta}}{\Gamma(\beta)}t^{\beta-1}\exp(-\alpha t).
\end{equation}

In this case the functional becomes
\begin{equation}
  \int_{0}^{\infty}(f_{t}'')^{2}dt=
    \sum_{i=1}^{5}\frac{W_{i}\alpha_{i}\Gamma(2\beta-i)}{2^{(2\beta-i)}(\Gamma(\beta))^{2}},
  \label{Eq:hopt2}
\end{equation}
where
\[
  W_{1}=A^{2},\;\;W_{2}=2AB,\;\;W_{3}=B^{2}+2AC,\;\;W_{4}=2BC,\;\;W_{5}=C^{2},
\]
and
\[
  A=\alpha^{2},\;\;B=-2\alpha(\beta-1),\;\;C=(\beta-1)(\beta-2).
\]

From Eq. (\ref{Eq:hopt2}), apparently, in order to get a nonzero
bandwidth, we has constrained $\beta$ to be at least equal to 3.
Using this constraint, we can obtain the estimates of the parameters
$\alpha$ and $\beta$ via the method of moments:
$\overline{t}=\widehat{E}(t)=\frac{1}{N}\sum_{i=1}^{N}t_{i}$ and
$\widehat{E}(t^{2})=\frac{1}{N}\sum_{i=1}^{N}t_{i}^{2}$ and the
sample standard deviation is
$\zeta=\sqrt{\widehat{E}(t^{2})-\overline{t}^{2}}$. The estimates
are $\widehat{\alpha}=\frac{\overline{t}}{\zeta^{2}}$ and
$\widehat{\beta}=\frac{\overline{t}^{2}}{\zeta^{2}}\geq 3$.

The kernel estimator can be easily generalized to the multivariate
case. Suppose we consider $X=[X_1,X_2,\dots,X_m]^\top$, let
$\overrightarrow{t}=[t_1,t_2,\dots,t_m]$, and
$\overrightarrow{s_i}=[s_{1i},s_{2i},\dots,s_{mi}]$,
$s_{ji}(j=1,2,\dots,m)$ is the first passage time for $X_j$. Then,
the multivariate kernel density estimator with kernel $K$ and window
width $h$ is defined by \cite{Silverman:1986}
\begin{equation}
  \widehat{f}(\overrightarrow{t})=\frac{1}{N}\sum_{i=1}^{N}K\left[h,(\overrightarrow{t}-\overrightarrow{s_i})\right],
\end{equation}
where
\begin{equation}
  K(\overrightarrow{t})=(2\pi h^2)^{-m/2}\exp\left(-\frac{1}{2h^2}\overrightarrow{t}^\top\overrightarrow{t}\right).
\end{equation}

And if we approximate the true density $f$ as a unit $m$-variate
normal density, then the optimal bandwidth $h_{opt}$ is
\cite{Silverman:1986}
\begin{equation}
  h_{opt}=\frac{1}{N^{1/(m+4)}}\left[\frac{4}{2m+1}\right]^{1/(m+4)}.
  \label{estamate:hopt:multi}
\end{equation}

\section{Algorithms}
\label{Methodology}

In section \ref{Model}, we have built up a multivariate
jump-diffusion model as describe in Eq. (\ref{JDP:multi}), and its
first passage time distribution was also obtained in section
\ref{subsection:FPTD}. In this section, we will discuss how to
simulate the multivariate jump-diffusion process efficiently via
Monte Carlo method. In conventional Monte Carlo method, the
simulation is very straightforward, we divide the time horizon
$[0,T]$ into $n$ small intervals $[0,t_1]$, $[t_1,t_2]$, $\cdots$,
$[t_{n-1},T]$ and in each Monte Carlo run, we need to calculate the
value of $X_i$ at each discretized time $t$. We should mention that
in order to exclude discretization bias, the number $n$ must be
large enough. It is obvious that this conventional method is very
time-consuming.

Recently, Atiya and Metwally \cite{Metwally:2002,Atiya:2005} have
proposed two fast Monte Carlo type methods, which are about 10-30
times faster than the conventional Monte Carlo approach. We called
them uniform sampling (UNIF) method, which involves sampling using
uniform distribution, and inverse Gaussian density sampling (IG)
method, which uses inverse Gaussian density method for sampling.

In this article, we mainly focus on the uniform sampling (UNIF)
method and extend it to the multivariate jump-diffusion process. The
major improvement of UNIF method is that it only evaluate $X_i$ at
generated jump instants and between each two jumps the process is a
Brownian bridge, so we just consider the probability of $X_i$
crossing the boundary level in $(T_{j-1},T_j)$ instead of evaluating
$X_i$ at each discretized time $t$. More exactly, we assume that the
values of $X_i(T_{j-1}^+)$ and $X_i(T_j^-)$ are known as two end
points of Brownian bridge, we generate a variable $s_i$ with uniform
distribution and by using Eq. (\ref{BM:default}) to see whether
$X_i(s_i)$ is smaller than the threshold level, if it defaults, then
we have successfully generated a first passage time $s_i$ and can
neglect the other intervals and perform another Monte Carlo run.

In \cite{Metwally:2002,Atiya:2005}, the jump-diffusion process is
involved in a univariate model. However, our model is based on
multivariate process in which $X_i$ are correlated as described in
Eq. (\ref{JDP:multi}), so we need to consider several points as
follows:
\begin{enumerate}
  \item In this article, as a first step, we assume that the arrival rate
$\lambda$ for the Poisson jump process and the distribution of
$(T_j-T_{j-1})$ are the same for each $X_i$. As for jump-size, we
should generate them as given distribution, and it can be different
to reflect the different jump process for each $X_i$.
  \item At present, we use exponential distribution (mean value $\mu_T$) for
$(T_j-T_{j-1})$ and normal distribution (mean value $\mu_J$ and
standard deviation $\sigma_J$) for the jump-size. Of course, we can
use any distribution as desired.
  \item If we consider $m$ processes, i.e., $X=[X_1,X_2,\dots,X_m]^\top$,
then we need an array \texttt{IsDefault} (whose size is $m$) to
indicate whether process $X_i$ has crossed the threshold in this
Monte Carlo run. If $X_i$ has crossed, then we set
\texttt{IsDefault}$(i)=1$, and will not evaluate it during this
Monte Carlo run.
\end{enumerate}

Next, we will give a description of our algorithm, based on a
multivariate extension of the algorithms proposed in
\cite{Metwally:2002,Atiya:2005}.

\subsection{Uniform sampling method}

Let us consider $m$ processes in the given time horizon $[0,T]$, as
described above, we have generated the jump instant $T_{j}$ by
generating interjump times $(T_{j}-T_{j-1})$, besides we set
\texttt{IsDefault}$(i)=0(i=1,2,\cdots,m)$ at first.

From Eq. (\ref{JDP:one}), we can see that,
\begin{enumerate}
  \item If jump doesn't occur, the diffusion follows a standard Brownian
motion, $W_{i}(T)\sim N(0,T)$, so interjump size
$(X_i(T_{j}^{-})-X_i(T_{j-1}^{+}))$ follows a normal distribution of
mean $\mu_i(T_{j}-T_{j-1})$ and standard deviation
$\sigma_i\sqrt{T_{j}-T_{j-1}}$. After extend if necessary, we get
\[
X_i(T_{j}^{-})\sim X_i(T_{j-1}^{+})+\mu_i(T_{j}-T_{j-1})+
\sum_{k=1}^{m}\sigma_{ik}N(0,T_{j}-T_{j-1}),
\]
and the initial state is $X_i(0)=X_i(T_{0}^{+})$.
  \item If jump occurs, the jump-size and direction of $Z_i(T_{j})$ are not
fixed either. We simulate the jump-size by a normal distribution,
and of course we may generate it according to other distribution.
Then we can compute the postjump value:
\[
  X_i(T_{j}^{+})=X_i(T_{j}^{-})+Z_i(T_{j}).
\]
\end{enumerate}

After generating beforejump and postjump value $X_i(T_{j}^{-})$ and
$X_i(T_{j}^{+})$ ($j=1,\cdots,M$, $M$ is the total number of jumps
for all the processes $X_i$), we can compute $P_{ij}$ according to
Eq. (\ref{BM:default}). To recur the first passage time density
(FPTD) $f_i(t)$, we need to consider three conditions for each $X_i$
that is still above the threshold:
\begin{enumerate}
  \item \textbf{First passage happens inside the interval.} We know if
$X_i(T_{j-1}^{+})>D_i(t)$, $X_i(T_{j}^{-})<D_i(t)$, then the first
passage happened in the time interval $[T_{j-1},T_{j}]$. To judge
when the first passage happen, first we compute the probability
$P_{ij}$ of $X_{i}$ always above the threshold according to Eq.
(\ref{BM:default}), then we generate $s_i$ as
$s_i=b_{ij}u_i+T_{j-1}$, where
$b_{ij}=\frac{T_{j}-T_{j-1}}{1-P_{ij}}$, and $u_i$ is a uniform
random number in $[0,1]$. If $s_i$ also belongs to interval
$[T_{j-1},T_{j}]$, then the first passage time occurred in this
interval, where $s_i$ is the first passage time and now we set
\texttt{IsDefault}$(i)=1$ to indicate process $X_i$ has crossed the
critical level. In this condition, we can get conditional interjump
first passage time density of that specific interval by
Eq.(\ref{FPTD:condition}). To get the density of whole interval
$[0,T]$, we have
$\widehat{f}_{i,n}(t)=\left(\frac{T_{j}-T_{j-1}}{1-P_{ij}}\right)g_{ij}(s_i)*K(h_{opt},t-s_i)$,
where $n$ is the iteration number of Monte Carlo cycle.
  \item \textbf{First passage doesn't happen in this interval.} If $s_i$
doesn't belong to interval $[T_{j-1},T_{j}]$, then the first passage
time has not yet occur in this interval.
  \item \textbf{First passage happens in the right boundary of interval.} If
$X_i(T_{j}^{+})<D_i(t)$, $X_i(T_{j}^{-})>D_i(t)$, which follow the
definition in Eq. (\ref{index_first_jump}), then obviously $T_{I_i}$
(set $I_i=j$) is the first passage time. Evaluate the density
function using kernel function
$\widehat{f}_{i,n}(t)=K(h_{opt},t-T_{I_i})$, and set
\texttt{IsDefault}$(i)=1$.
\end{enumerate}

Then we increase $j$ and examine the next interval and judge these
three conditions for each non-crossing process $X_i$ again. For each
Monte Carlo run, if we make a rough assumption that the probability
of $X_i$ crossing the threshold is not correlated, then we can
obtain the multivariate FPTD as
$\widehat{f}_{n}(\overrightarrow{t})=\Pi_{i=1}^{m}\left(\frac{T_{j}-T_{j-1}}{1-P_{ij}}\right)g_{ij}(s_i)*K(h_{opt},\overrightarrow{t}-\overrightarrow{s})$.

After running $N$ times of Monte Carlo cycle, we get the
one-dimensional FPTD of $X_i$ as
$\widehat{f}_{i}(t)=\frac{1}{N}\sum_{n=1}^{N}\widehat{f}_{i,n}(t)$,
and multivariate FPTD as
$\widehat{f}(\overrightarrow{t})=\frac{1}{N}\sum_{n=1}^{N}\widehat{f}_{n}(\overrightarrow{t})$

\section{Simulation results}
\label{Simulations}

In this section, as a demonstration, we will test the multivariate
UNIF method on two-dimensional case. In order to check the efficient
and validity of the UNIF method, we use three examples with
different arrival rate $\lambda=1,3,8$ for the Poisson jump process
to judge the efficiency of our algorithms. The parameters are as
follows
\begin{eqnarray*}
  && X_0=[0,0]^\top,\;
  D(t)=[\ln(0.9)-0.002t,\ln(0.95)-0.012t]^\top\\
  && \mu=[-0.002,-0.012]^\top,\;
  \sigma=\left[\begin{tabular}{cc}
     $0.2$ & $0.0$\\
     $0.0$ & $0.2$
   \end{tabular}\right],\\
  && \mu_Z=[0,0]^\top,\;
  \sigma_Z=[0.2,0.12]^\top.
\end{eqnarray*}
where $X_0$ is the starting value for the process, $D_(t)$ is the
threshold, $\mu$ is the constant instantaneous drift, $\sigma$
represents the Brownian motion, and $\mu_Z$ and $\sigma_Z$ are the
mean and standard deviations, respectively, of the jump sizes.

\begin{figure}[hbtp]
  \centering
  \includegraphics[width=10.5cm]{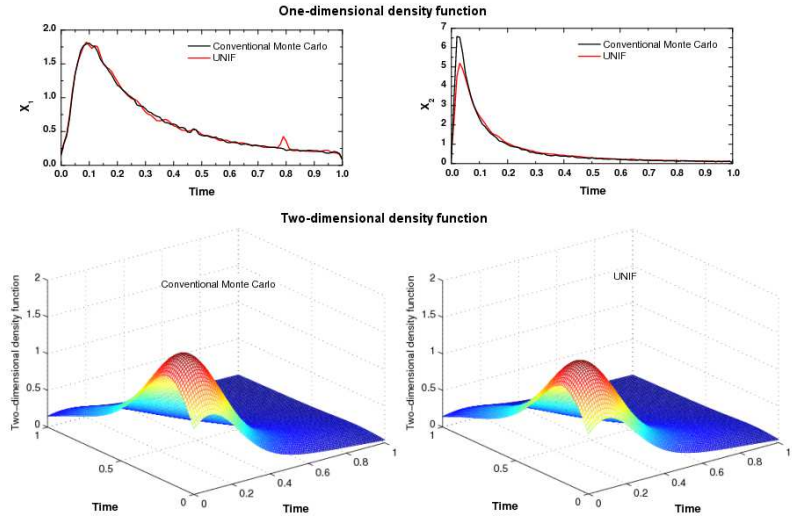}
  \caption{Example 1 ($\lambda=1$): One-dimensional (top) and two-dimensional
density function (bottom) estimate using $100,000$ iterations for
UNIF and conventional Monte Carlo approaches. The discretization
size of time horizon is $\Delta=0.0002$ for conventional Monte Carlo
method.}
  \label{Fig:example1}
\end{figure}

\begin{figure}[hbtp]
  \centering
  \includegraphics[width=10.5cm]{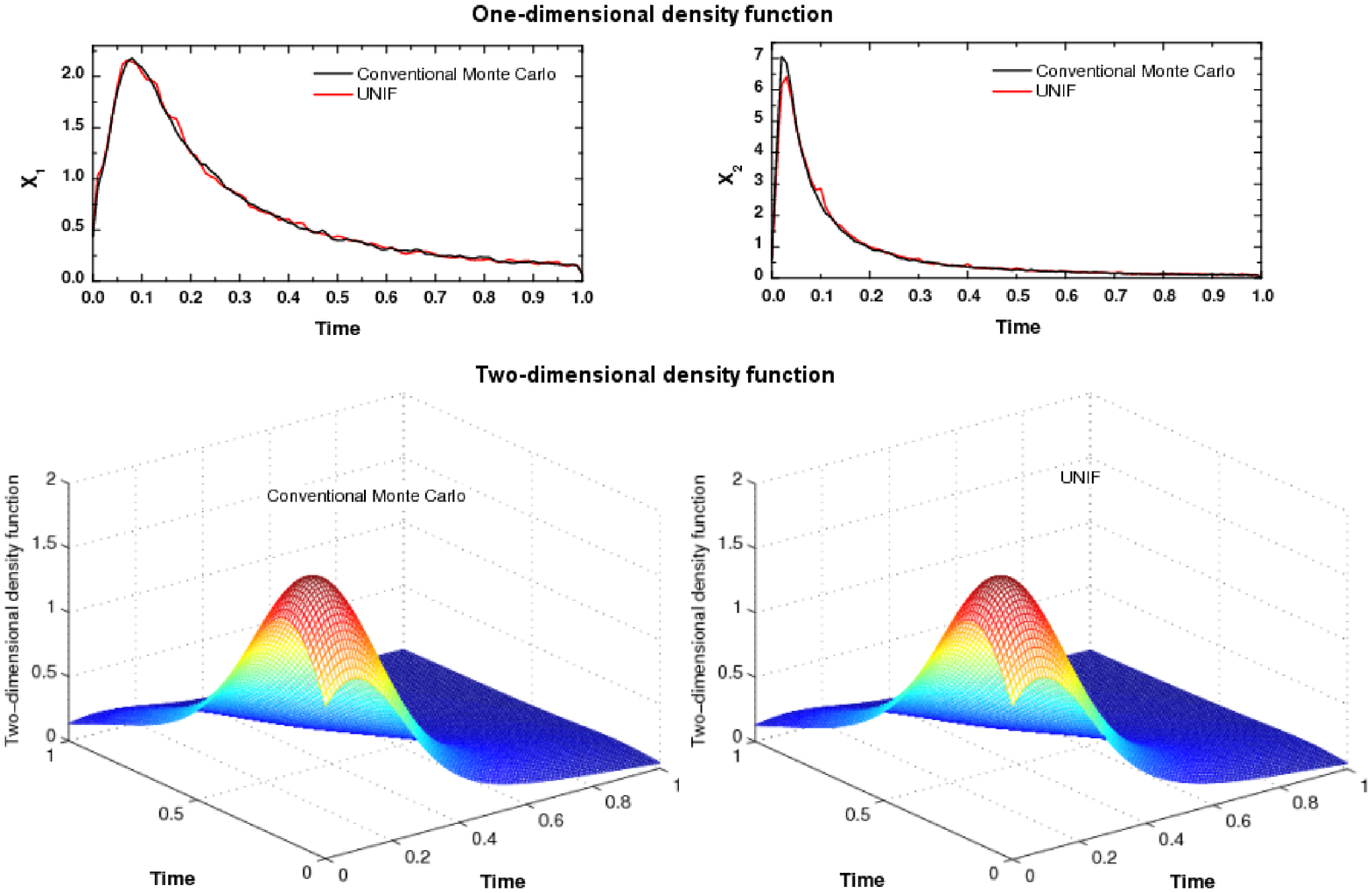}
  \caption{Example 2 ($\lambda=3$): One-dimensional (top) and two-dimensional
density function (bottom) estimate using $100,000$ iterations for
UNIF and conventional Monte Carlo approaches. The discretization
size of time horizon is $\Delta=0.0002$ for conventional Monte Carlo
method.}
  \label{Fig:example2}
\end{figure}

\begin{figure}[hbtp]
  \centering
  \includegraphics[width=10.5cm]{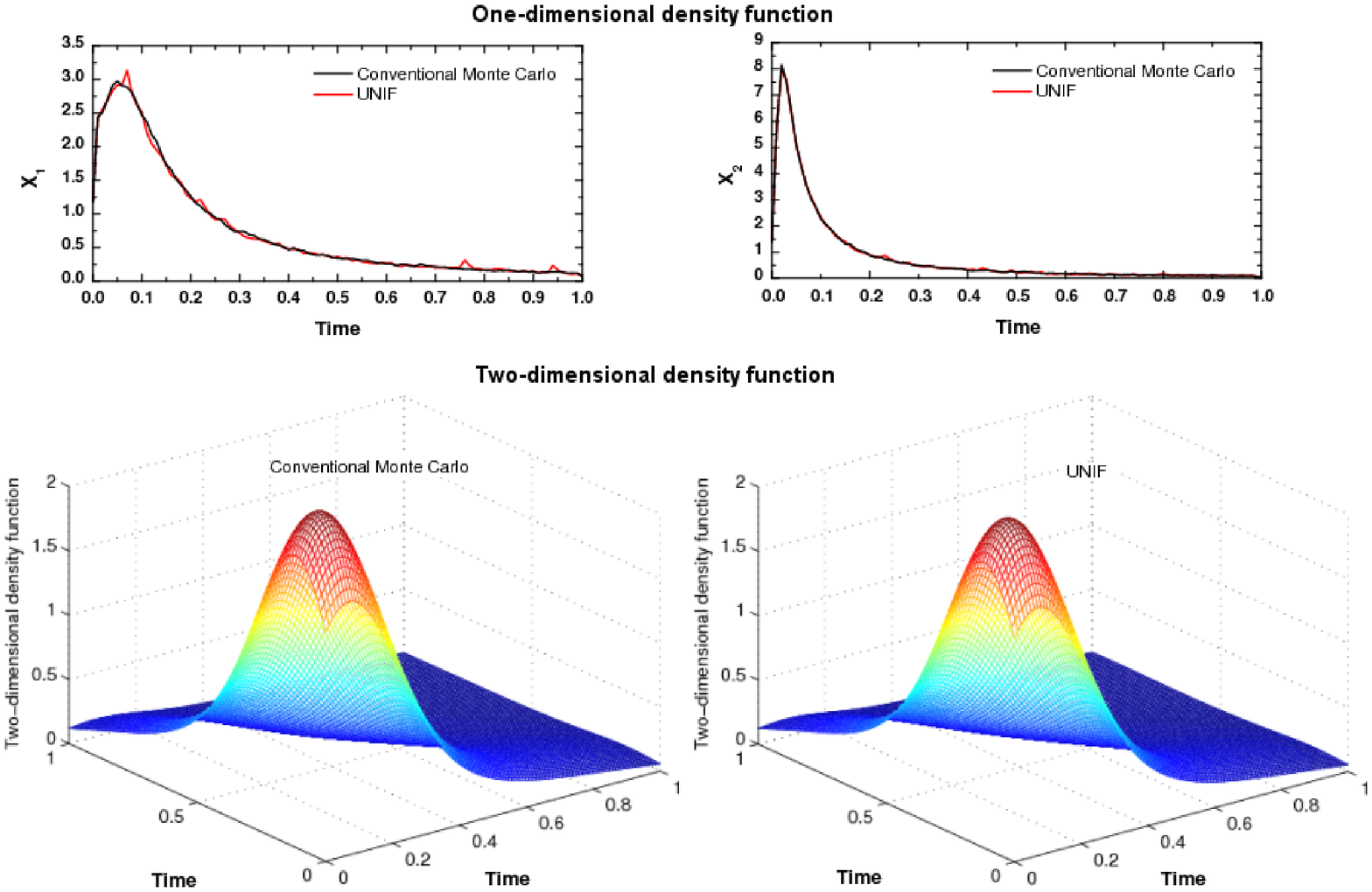}
  \caption{Example 3 ($\lambda=8$): One-dimensional (top) and two-dimensional
density function (bottom) estimate using $100,000$ iterations for
UNIF and conventional Monte Carlo approaches. The discretization
size of time horizon is $\Delta=0.0002$ for conventional Monte Carlo
method.}
  \label{Fig:example3}
\end{figure}

The simulation was carried out with total Monte Carlo runs
$N=500,000$ in horizon $[0,1]$. Moreover, we have also carried out
conventional Monte Carlo simulation with the same parameters, the
estimated density functions are displayed in Fig.
\ref{Fig:example1}-\ref{Fig:example3}. All the simulations were
carried out on a 2.4 GHz AMD Opteron(tm) Processor. The optimal
bandwidth and CPU time are described in Table \ref{Tab:hopt}.

From Fig. \ref{Fig:example1}-\ref{Fig:example3}, we can see that the
multivariate UNIF method gives similar density function as the
conventional one, that check the validity of our algorithms. An
interesting phenomenon is that increasing $\lambda$ will affect FPTD
of $X_1$ a lot whose probability of crossing the threshold is low in
the interval $[0,T]$.

\begin{table}[hbtp]
  \centering
  \caption{The optimal bandwidth $h_{opt}$, and CPU time per Monte Carlo run
of the simulations. The first $h_{opt}$ is for $X_1$ and the second
for $X_2$. All the simulations were performed with Monte Carlo runs
$N=100,000$, besides, for conventional Monte Carlo (CMC) method, the
discretization size of time horizon is $\Delta=0.0002$.}
  \label{Tab:hopt}
  \begin{tabular}{l|lccc}
    \hline
    & & \multicolumn{2}{c}{Optimal bandwidth} & CPU time\\
    & & $X_1$ & $X_2$ &\\
    \hline
    Example 1 & CMC  & 0.012664 & 0.006943 & 0.286642\\
              & UNIF & 0.016030 & 0.013880 & 0.000527\\
    Example 2 & CMC  & 0.011157 & 0.006582 & 0.284554\\
              & UNIF & 0.012249 & 0.009443 & 0.000731\\
    Example 3 & CMC  & 0.008894 & 0.005921 & 0.299156\\
              & UNIF & 0.009117 & 0.006542 & 0.001222\\
    \hline
  \end{tabular}
\end{table}

Undoubtedly, from Table \ref{Tab:hopt}, one can easily realizes that
the multivariate UNIF approach is much more efficient than the
conventional one, which establish it as a fast methodology for
practical applications.

\section{Conclusion}
\label{Conclusion}

In summary, we have studied the first passage time problem in the
context of multivariate jump-diffusion processes. We have extended
the fast Monte Carlo-type numerical method - the UNIF method -- to
multiple processes. From our simulation results, we can see that the
multivariate UNIF approach is much more efficient than the
conventional Monte Carlo method, which illustrates that the
developed methodology can provide an efficient tool for further
practical applications, such as the analysis of default correlation
and predicting barrier options in finance.

\noindent {\it Acknowledgements}. We would like to thank NSERC for
its support.

\medskip

Received Month Year; Sep.29, 2006.


\begin{thebibliography}{99}

\bibitem{Zhou:2001:jump}
  C. Zhou,
  \emph{The Term Structure of Credit Spreads with Jump Risk},
  Journal of Banking and Finance 25 (2001), 2015--2040.

\bibitem{Gorov:1985}
  G. Gorov, Ya. Kogan and N. Paradizov,
  \emph{Jump diffusion approximation in single-server systems with interruption of service
and variable rate of arrival of calls},
  Avtomatika i Telemekhanika, 6 (1985), 44--51.

\bibitem{Zhu:1999}
  S. C. Zhu,
  \emph{Stochastic jump-diffusion process for computing medial
axes in Markov random fields},
  IEEE Trans. Pattern Analysis and Machine Intelligence, 21 (1999), 1158--1169.

\bibitem{Miller:1997}
  M. Miller, U. Grenander, J. O'Sullivan and D. Snyder,
  \emph{Automatic target recognition organized via jump-diffusion algorithm},
  IEEE Trans. Image Processing, 6 (1997), 157--174.

\bibitem{Kou:2003}
  S.G. Kou, H. Wang,
  \emph{First passage times of a jump diffusion process},
  Adv. Appl. Probab. 35 (2003), 504--531.

\bibitem{Blake:1973}
  I. Blake, W. Lindsey,
  \emph{Level-crossing problems for random processes},
  IEEE Trans. Inform. Theory IT-19 (1973), 295--315.

\bibitem{Platen:2003}
  P.E. Kloeden, E. Platen, H. Schurz,
  ``Numerical Solution of SDE Through Computer Experiments'', Third Revised Edition,
  Springer, Germany, 2003.

\bibitem{Metwally:2002}
  S. Metwally, A. Atiya,
  \emph{Using brownian bridge for fast simulation of jump-diffusion processes and barrier options},
  The Journal of Derivatives 10 (2002), 43--54.

\bibitem{Atiya:2005}
  A.F. Atiya, S.A.K. Metwally,
  \emph{Efficient Estimation of First Passage Time Density Function for Jump-Diffusion Processes},
  SIAM Journal on Scientific Computing 26 (2005), 1760--1775.

\bibitem{Duffie:2000}
  D. Duffie, J. Pan, K. Singleton,
  \emph{Transform Analysis and Option Pricing for Affine Jump-Diffusions},
  Econometrica 68 (2000), 1343--1376.

\bibitem{Feller:1968}
  W. Feller,
  ``An Introduction to Probability Theory and Its Applications'', Vol. 1,
  Wiley, New York, 1968.

\bibitem{Rogers:1994}
  L.C.G. Rogers and D. Williams,
  ``Diffusions, Markov Processes and Martingales'', Vol. 1,
  Wiley, New York, 1994.

\bibitem{Karatzas:1991}
  I. Karatzas and S. Shreve,
  ``Brownian Motion and Stochastic Calculus'',
  Springer-Verlag, New York, 1991.

\bibitem{Costa:1995}
  O. Costa, M. Fragoso,
  \emph{Discrete time LQ-optimal control problems for infinite Markov jump parameter systems},
  IEEE Trans. Automat. Control 40 (1995), 2076--2088.

\bibitem{Silverman:1986}
  B.W. Silverman,
  ``Density Estimation for Statistics and Data Analysis'',
  Chapman \& Hall, London, 1986.

\bibitem{Devroye:1986}
  L. Devroye,
  ``Non-uniform Random Variate Generation'',
  Springer-Verlag, New York, 1986.

\end{thebibliography}
\end{document}